\documentclass[aps,pre,preprint,amsfonts,amsmath,amssymb,showpacs,titlepage]{revtex4-1}
\usepackage{graphicx, mathtools, hyperref, comment}
\usepackage[caption=false]{subfig}
\usepackage{color}

\newcommand*{\avg}[1]{\langle #1 \rangle} 
\renewcommand*{\d}[2]{\frac{\mathrm{d} #1}{\mathrm{d} #2}} 

\begin{document}

\title{Finite-time Lyapunov exponents in chaotic Bose-Hubbard chains}

\author{R. A. Kidd}
\author{M. K. Olsen}
\author{J. F. Corney}
\affiliation{School of Mathematics and Physics, University of Queensland, Queensland 4072, Australia.}

\date{\today}

\begin{abstract}

Many-site Bose-Hubbard lattices display complex semiclassical dynamics, with both chaotic and regular features. We have characterised chaos in the semiclassical dynamics of short Bose-Hubbard chains using both stroboscopic phase space projections and finite-time Lyapunov exponents. We found that chaos was present for intermediate collisional nonlinearity in the open trimer and quatramer systems, with soft chaos and Kolmogoroff-Arnold-Moser islands evident. We have found that the finite-time Lyapunov exponents are consistent with stroboscopic maps for the prediction of chaos in these small systems. This gives us confidence that  the finite-time Lyapunov exponents will be a useful tool for the characterisation of chaos in larger systems, where meaningful phase-space projections are not possible and the dimensionality of the problem can make the standard methods intractable.

\end{abstract}

\pacs{05.45.Mt, 05.45.Pq, 03.65.Sq, 03.75.Lm}

\maketitle

\section{Introduction}
\label{sec:intro}

Ultra-cold bosons in optical lattices offer a well-researched and highly-tuneable model system for exploring complex quantum many-body physics~\cite{Bloch2005}. Lattices of varying geometry can be generated by interfering laser beams to form a periodic potential with configurable period. In the limit that there is only one atomic mode per lattice site, these systems may be investigated using the Bose-Hubbard model~\cite{BHmodel,Jaksch}.
This model assumes that the condensate bosons are tightly-bound to the lattice sites with a single mode at each site and can only move via tunnelling to adjacent modes~\cite{Jaksch}.
The many-body dynamics of this model are particularly intriguing in regimes that show chaos in the semiclassical limit.  Such quantum chaos brings into sharp focus the issue of the classical-quantum transition, and can be thoroughly investigated in ultracold lattice systems.  The tunability of the parameters allows a full exploration of both regular and chaotic regions in phase space, and the choice of atom number controls how well the semiclassical description applies  (the effective Planck constant is inversely proportional to the total number).  The nonlinearity necessary for chaos arises from interatomic collisions~\cite{Dalfovo1999}, whose strength, relative to the linear tunnelling, is experimentally adjustable.

In this work we explore signatures of chaos in a semiclassical treatment of three- and four-site Bose Hubbard chains using finite-time Lyapunov exponents (FTLE)~\cite{FTLE}. We evaluate these and compare them with stroboscopic sections~\cite{Poincare} as a measure of characterising chaos.
Our analysis builds on and extends prior work on the modulated double-well potential~\cite{Joeldrive,Salmond2002,Boukobza2010}, the three-site chain~\cite{Buonsante2003,Franzosi2001,Mossman2006}, and ring configurions~\cite{Chiancathermal}.  In particular, we show that finite-time Lyapunov exponents are an effective and easily computable measure of semiclassical chaos for multimode systems, where the dimensionality of the phase-space can make it difficult to produce meaningful stroboscopic sections.
Although Lyapunov exponents and stroboscopic maps are in many ways complementary tools, the utility of stroboscopic maps does depend on choosing an appropriate section through multi-dimensional phase space~\cite{Mukherjee}, while Lyapunov exponents provide a rigorous determination of chaos for a given initial condition, even for large chains. Finite-time Lyapunov exponents have previously been used to analyse chaos on the trimer ring~\cite{Chiancathermal}. For lattice models with up to 41 sites~\cite{Cassidy2009}, they have been used over short times to investigate the relation between chaos and thermalisation.

In the semiclassical approximation, the expectation values of quantum observables are assumed to obey classical dynamical equations, obtained through factorising higher-order moments and usually valid for macroscopic particle number. Chaotic behaviour requires sufficient complexity or asymmetry to break the integrability of the system~\cite{Wimberger2014}.
With the two-site system (dimer) the semi-classical dynamics are integrable~\cite{JoelBH}, with this no longer being the case when driving is introduced~\cite{Joeldrive,Salmond2002,Boukobza2010}. In the absence of driving, the simplest Bose Hubbard model that exhibits classical chaos is the trimer.
Chaos in the open semiclassical trimer has been characterised through stroboscopic sections in a system where the central well depth and tunnelling were varied~\cite{Buonsante2003}, performing an analysis of the dynamical fixed points and the stability.  The appearance of chaos in the semiclassical system has been correlated with various features in the quantum description, including the structure of eigenstates~\cite{Mossman2006}, Husimi Q distributions~\cite{Trimborn2009} and the level spacing in the  energy spectrum~\cite{Tikhonenkov2013}.  The four-site  model is referred to as a quatramer and, being more complex than the three-site system, is also expected to exhibit chaotic behaviour due to non-integrability.

\section{Bose-Hubbard model}
\label{sec:BH}

The Bose-Hubbard model with tunnel-coupling and collisional interactions is described by the Hamiltonian
\begin{equation}
  \hat H = \hbar\chi \sum_{j} \hat a_j^\dagger \hat a_j^\dagger \hat a_j \hat a_j - \hbar J \sum_{\avg{jk}}  \hat a_j^\dagger \hat a_k
  \label{eq:Hamiltonian}
\end{equation}
where the $\hat{a}_{j}$ are the bosonic annihilation operators for each site, $\chi$ represents the collisional interaction strength, and $J$ is the rate of the inter-site coupling. The nearest-neighbour notation, $\avg{jk}$, is used to indicate summation over neighbouring pairs of  wells. The semiclassical equations of motion can be obtained simply by replacing the bosonic operators with complex numbers in the Heisenberg equations of motion and and assuming that all moments factorise. This gives us
\begin{equation}
  \d{\alpha_j}{t} = -2i\chi |\alpha_j|^2 \alpha_j + iJ \sum_{k=j\pm 1}  \alpha_k,
  \label{eq:ODE}
\end{equation}
where the $\alpha_{j}$ are classical variables representing atomic amplitudes. As has often been shown~\cite{JoelBH,Chiancathermal}, this correspondence is only valid for short evolution times, but is generally considered to be more valid as the occupation numbers in the wells increases~\cite{Dalfovo1999}.

Eq.~\eqref{eq:ODE} can be solved analytically in some limiting cases, but is readily solved numerically for a moderate number of sites, with the computational complexity only increasing linearly in the number of wells. Although we cannot calculate any quantum correlations using the above equation, it can be integrated over the long evolution times required to evaluate the FTLEs.

\section{Special unitary groups}
\label{sec:groups}

The dynamical group of a Bose-Hubbard trimer is SU(3) (Special Unitary 3) and has been used by various researchers~\cite{Nemoto2000, Franzosi2001, Viscondi2011, Jason2012}. In general, SU($M$) is the dynamical group for the $M$-site  model~\cite{Davidson2015}. The SU($M$) generators lead to a convenient set of phase space variables, as the expectation values of the operators are all real and represent observable quantities. Although the SU(M) groups are usually expressed in terms of operator functions, we may define their semiclassical equivalents for the purposes of our investigation.

The eight generators of SU(3) written in terms of the ladder operators are
\begin{align}
  \hat x_1 &= \hat{a}_1^\dagger\hat{a}_1 - \hat{a}_2^\dagger\hat{a}_2,  &
  \hat x_2 &= \frac{1}{3} \left(\hat{a}_1^\dagger\hat{a}_1 + \hat{a}_2^\dagger\hat{a}_2 - 2\hat{a}_3^\dagger\hat{a}_3\right), \nonumber \\
  \hat y_j &= i \left( \hat{a}_j^\dagger\hat{a}_k - \hat{a}_k^\dagger\hat{a}_j \right), &
  \hat z_j &= \hat{a}_j^\dagger\hat{a}_k + \hat{a}_k^\dagger\hat{a}_j,
  \label{eq:SU(3)}
\end{align}
and consist of occupation, current and coherence operators. The occupation operators, $\hat x_1$ and $\hat x_2$, represent the two-site and three-site particle number differences, respectively. The three current operators, $\hat y_j$, represent the particle currents between sites $j=1,2,3$ and subsequent sites $k = j+1 \mod{3}$. The three coherence operators, $\hat z_j$, represent the coherence of site $j$ with respect to adjoining site $k$. We define their classical equivalents as
{\small
\begin{align}
  x_1 &= \alpha_{1}^{\ast}\alpha_{1} - \alpha_{2}^{\ast}\alpha_{2}, &
  x_2 &= \frac{1}{3} \left(\alpha_{1}^{\ast}\alpha_{1}+\alpha_{2}^{\ast}\alpha_{2}-2\alpha_{3}^{\ast}\alpha_{3}\right), \nonumber \\
  y_j &= i \left( \alpha_{j}^{\ast}\alpha_{k}-\alpha_{k}^{\ast}\alpha_{j}\right), &
  z_j &= \alpha_{j}^{\ast}\alpha_{k}+\alpha_{k}^{\ast}\alpha_{j}.
  \label{eq:SU(3)class}
\end{align}}

The SU(4) operators consist of the SU(3) operators, an additional occupation operator, given in Eq.~\eqref{SU(4)}, three additional current operators, and three additional coherence operators. The additional current and occupation operators are defined by extension from the definitions of $\hat{y}_j$ and $\hat{z}_j$, for $i=4$ and $k=1,2,3$. The additional occupation operator is
\begin{equation}
  \hat{x}_3 = \frac{1}{6} \left(- 3\hat{a}_4^{\dagger}\hat{a}_{4} + \sum_{j=1}^3\hat{a}_j^\dagger\hat{a}_j \right).
  \label{SU(4)}
\end{equation}

\section{Signatures of chaos}
\label{sec:signatures}

For systems with many variables, the phase-space dynamics can be difficult to visualise and analyse via the dynamics of the separate variables. To overcome this problem we will use stroboscopic phase space maps, constructed by plotting trajectories periodically in time. This allows us to visualise the distinction between invariant tori and chaotic behaviour. As the time-evolution of Hamiltonian systems is unique, the representative stroboscopic section must be similarly so~\cite{Wimberger2014}. Regions of local stability are represented in stroboscopic sections as closed loops, while regions of chaos are represented by points that appear to be stochastically distributed~\cite{Wimberger2014}, filling regions of the phase space. Therefore, stroboscopic sections readily show the transition between stable and chaotic dynamics.

Classical phase space stroboscopic sections of the open trimer~\cite{Mossman2006} are known to display soft chaos and Kolmogoroff-Arnold-Moser (KAM) islands~\cite{Gutzwiller1990}. The KAM theorem~\cite{Gutzwiller1990} states that certain invariant tori in the phase space of an integrable system will remain in the presence of non-resonant  perturbations. The intermingling of invariant tori and regions of chaos is referred to as `soft chaos'. Clusters of invariant tori are referred to as KAM islands~\cite{Wimberger2014}.
Trimer regimes with constant tunnel-coupling have been shown to exhibit two distinct regimes, one for low and one for high total system energy~\cite{Mossman2006}. Low energy regimes exhibit coupling between all three sites and display soft chaos and large KAM islands, with concentric tori centred on a stable fixed point~\cite{Mossman2006}. Another fixed point is present in the form of an unstable saddle point surrounded by chaotic regions, acting as the centre of a homoclinic tangle~\cite{Mossman2006}. For increasing energy, the KAM islands shrink and a large chaotic sea grows around them~\cite{Mossman2006}. High energy regimes exhibit frequent three-site coupling and decoupling and display large scale chaos with small, isolated KAM islands~\cite{Mossman2006}. The total system energy depends on the number of wells, their occupation, the tunneling rate, and the collisional interaction energy, so that these systems are in principle widely tunable.

The finite-time Lyapunov exponent~\cite{FTLE}, $\lambda$, defined below, provides a means of quantifying a system's dynamical sensitivity to initial conditions. The complex vector $\vec{Z}_0$ is the initial phase space point under consideration at $t=0$ and $\vec{Z}'_0$ is a slight perturbation from $\vec{Z}_0$. In the work presented here, the perturbations were three orders of magnitude larger than the numerical precision used. Positive Lyapunov exponents represent sensitivity to initial conditions, which is a signature of chaos.
\begin{equation}
  \lambda(t) = \lim_{\vec{Z}'_0 \to \vec{Z}_0}  \frac{1}{t} \ln{\frac{||\vec{Z}'(t)-\vec{Z}(t)||}{||\vec{Z}'(0)-\vec{Z}(0) ||}}.
  \label{eq:Lyapunov}
\end{equation}
There are various options for $\vec{Z}$, as long as we are consistent. We can, for example, define it in terms of the variables of Eq.~\eqref{eq:SU(3)class}, or their four-site equivalents. For the trimer, this gives
\begin{equation}
  \vec{Z} = \left(x_{1},x_{2},y_{1},y_{2},y_{3},z_{1},z_{2},z_{3} \right),
  \label{eq:ZSUtrimer}
\end{equation}
while an alternative definition in terms of the variables used in the equations of motion gives
\begin{equation}
  \vec{Z} = \left( \alpha_1, \alpha_2,\alpha_{3} \right).
  \label{eq:Zalpha}
\end{equation}

In a regular, non-chaotic system, the finite-time Lyapunov exponents will tend to decrease with time, as a small difference in initial conditions is expected to lead to trajectories that closely follow one another. Alternately, exponents that do not decrease with time indicate that there are exponentially divergent trajectories present.

\section{Results and discussion}
\label{sec:discuss}

We have numerically integrated Eq.~\eqref{eq:ODE} for both the trimer and quatramer configurations, for various values of the interaction strength $\chi$.  To construct the stroboscopic map,  we scan through initial conditions:
\begin{equation}
  \alpha_j(0) = \begin{cases} \sqrt{k},  & j<M \\ \sqrt{(M-1)(100-k)}, & j=M \end{cases},
  \label{eq:init1}
\end{equation}
where $k$ is an integer that ranges between 0 and 100.  Note that, in order to obtain reduce the visual complexity of the plots, we have only considered a subset of all possible initial conditions, namely, the relative phase is initially zero, and all wells except the end one contain the same number of particles.  With this choice, the total number $N$, in the $M$-site system given by
\begin{equation}
  N = \sum_{j=1}^M |\alpha_j|^2 = 100(M-1).
  \label{eq:init2}
\end{equation}
This choice of total number is made to facilitate comparison in a possible experimental scenario. The results are actually more general, since in a semiclassical analysis it is the nonlinearity, $\chi N$, rather than $\chi$ or $N$ separately, that determines the dynamical regime.

To generate the stroboscopic map, we plot points at time intervals  $\Delta t = \pi/2$, up to total time $t = 1000$, in the reduced phase space spanning $x_2$, $y_3$ and $z_3$. These variables represent the three-site occupation difference, the current between sites 2 and 3, and the coherence between sites 2 and 3, respectively. Out of the eight SU(3) operators, these were chosen because together they completely describe the state of the end well.  For the four-site system, we chose $x_3$, $y_6$ and $z_6$ from the SU(4) variables. These variables represent the four-site occupation difference and the current and coherence between sites 3 and 4.

As the canonical Lyapunov exponents are defined in the limit of large $t$~\cite{FTLE}, the validity of FTLEs increases with time. However, as the phase space is finite, the divergence of trajectories resulting from perturbed and unperturbed initial conditions reaches a maximum value for finite-time. In this paper, the FTLEs were calculated at a time large enough so that the chaotic trajectories diverged from the stable ones, and both approached distinct asymptotes.

\begin{figure}[ht]
  \centering
  \subfloat{\includegraphics[width=.4\columnwidth]{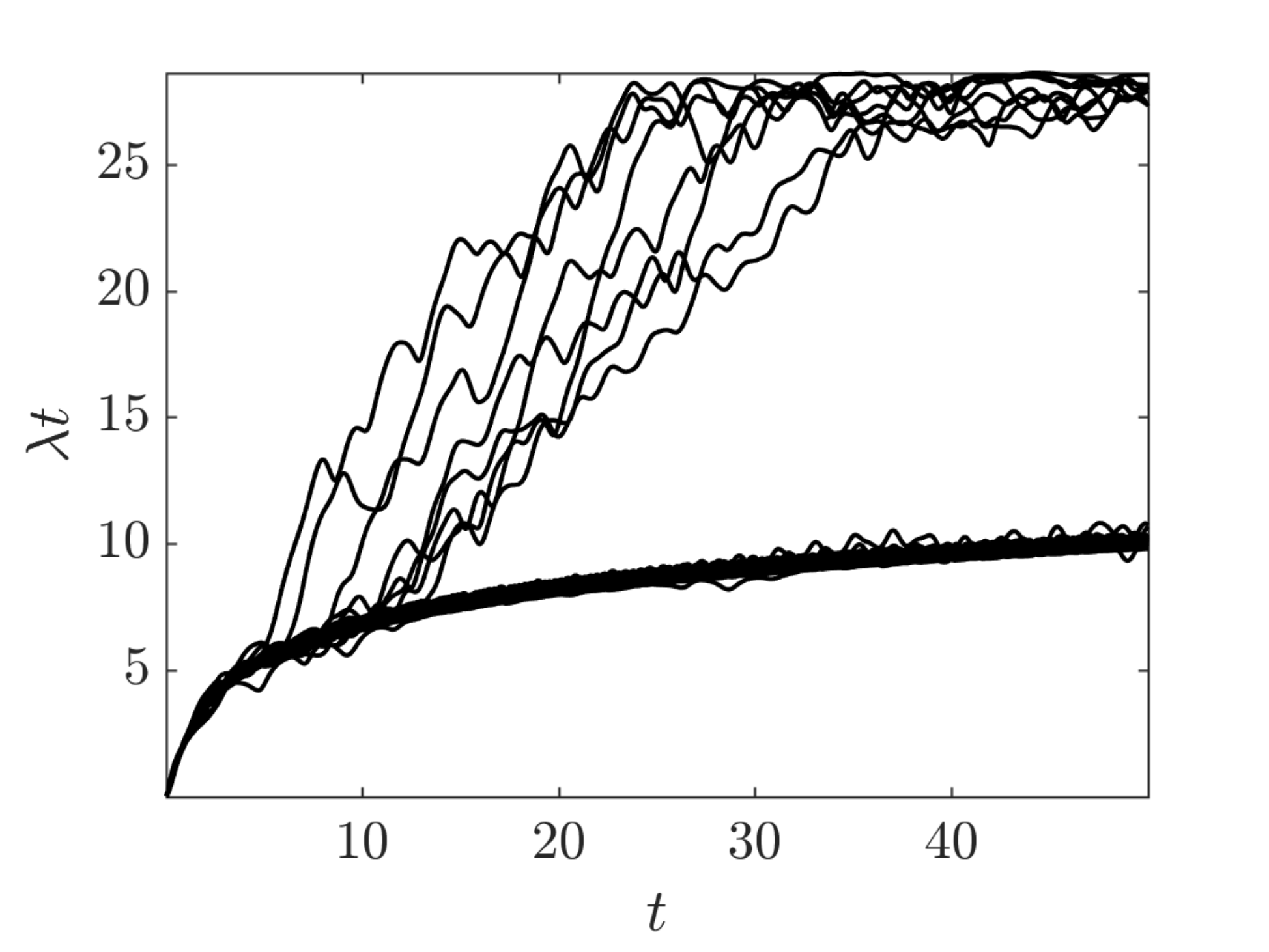} \label{fig:lyaptime1}}

  \subfloat{\includegraphics[width=.4\columnwidth]{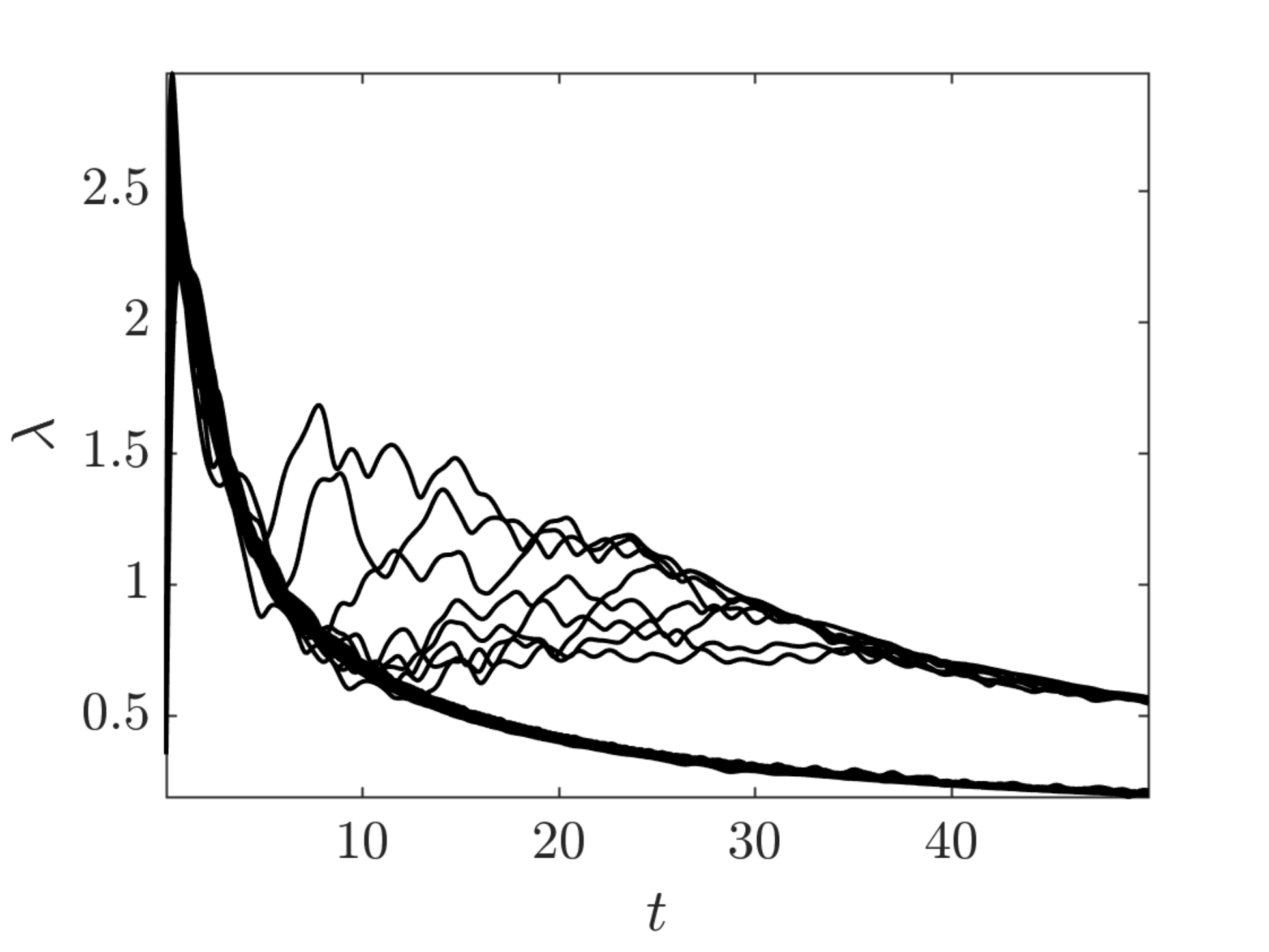}}

  \subfloat{\includegraphics[width=.45\columnwidth]{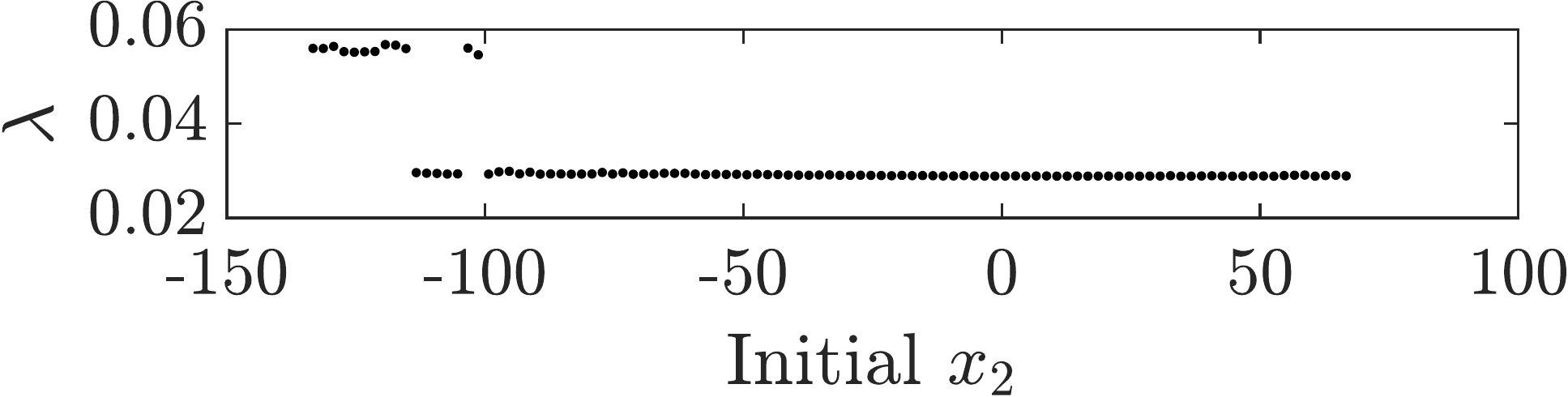}}

  \caption{Trimer FTLEs, $\lambda$, as a function of dimensionless time, $t$, for $N\chi/J=2$ and a perturbation of $\Delta\alpha \sim 10^{-6}$. The top plot shows the initial linear increase in $\lambda t$, an indicator of chaotic trajectories, compared to the logarithmic $\lambda t$ for stable trajectories. The middle plot shows the logarithms of the divergences of perturbed trajectories, $\lambda t$. The bottom plot shows FTLEs at $t=500$ as a function of initial $x_2$. This time is chosen as several trajectories diverged from the stable FTLE region and converged to the unstable region of relatively larger magnitude.}
  \label{fig:lyaptime}
\end{figure}

How the FTLEs work in practice is shown in Fig.~\ref{fig:lyaptime1}. We see that the plots of $\lambda t = \ln{\frac{||\vec{Z}(t) - \vec{Z}'(t)||}{||\vec{Z}_0 - \vec{Z}'_0||}}$ for stable trajectories increase logarithmically, indicating linear divergence of trajectories. The lines in Fig.~\ref{fig:lyaptime1} that initially increase linearly indicate exponential divergence of these perturbed trajectories. The levelling off that occurs after $t=20$ indicates that the divergence of trajectories is limited by the phase space boundary. After $t=20$, several trajectories diverged from the lower region and converged to the upper, indicating that their initial conditions were near the boundary between chaotic and regular regimes.

On a final note, we found that the FTLEs for the quatramer took longer to converge than for the trimer. This effect is explicable in terms of the increased phase space volume explored by the quatramer. We also note that the FTLEs that we display in this work are defined in terms of the site amplitudes, as in Eq.~\eqref{eq:Zalpha}. The FTLEs produced using the SU($M$) variables were consistent with these, but the amplitude method is more easily extensible to higher numbers of sites, as the number of SU($M$) variables increases quadratically with $M$.

\subsection{The open trimer}
\label{subsec:trimer}

\begin{figure}[ht]
  \centering
  \subfloat{\includegraphics[width=.45\columnwidth]{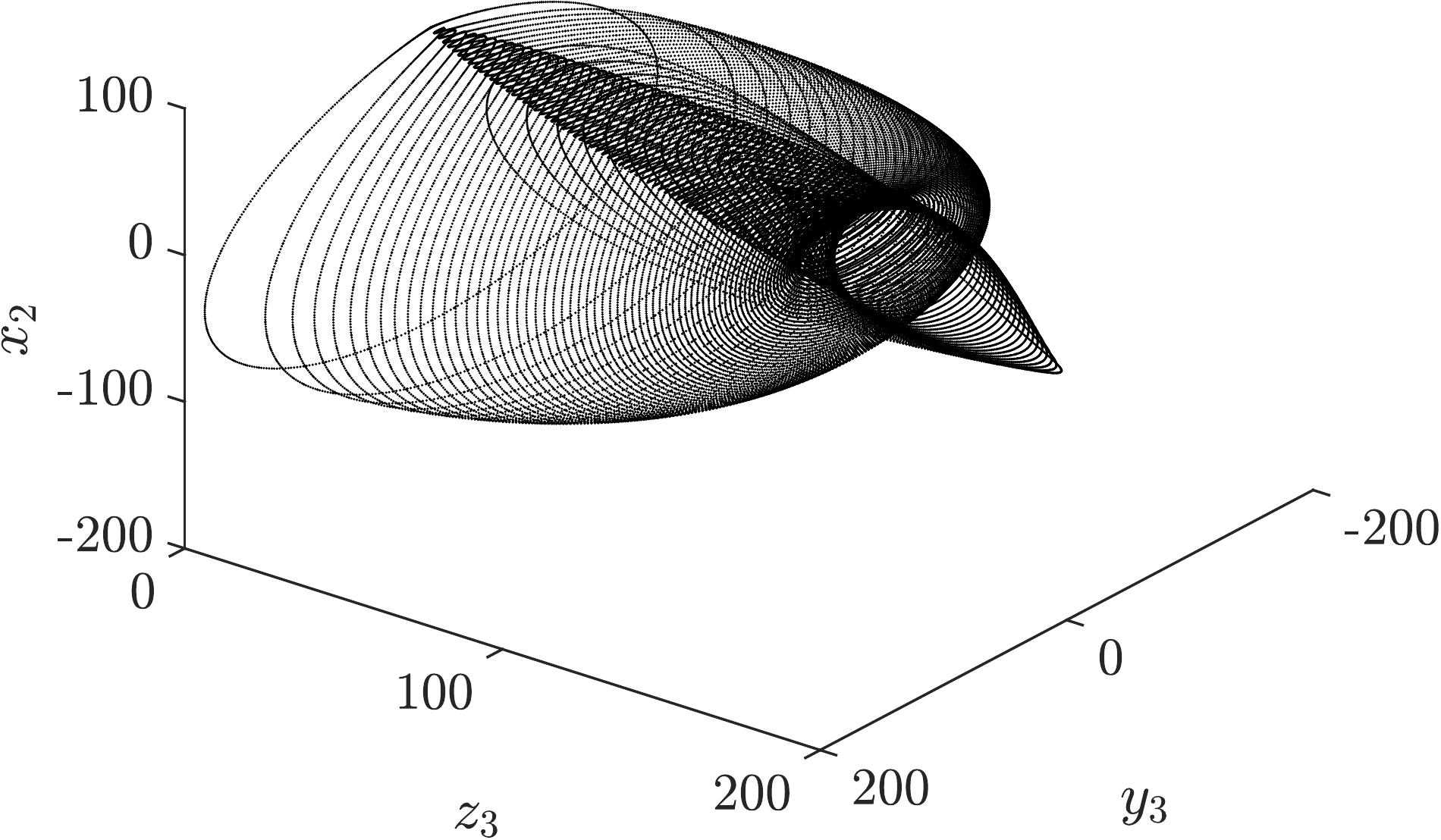}}

  \subfloat{\includegraphics[width=.45\columnwidth]{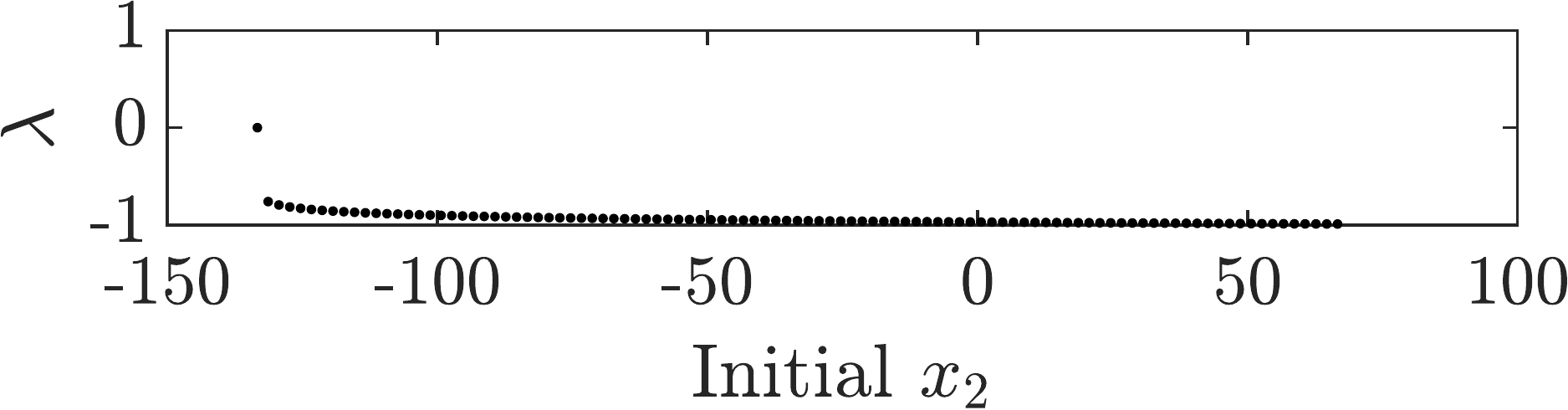}}
  \caption{Regular 3D trimer phase space section (above) for $N\chi/J=0$ and FTLEs (below) at $t=20$.}
  \label{fig:trimer1}
\end{figure}

We begin by investigating the FTLEs for the open trimer with $N\chi/J=0$, where the system is linear and hence chaotic behaviour is not expected.
As shown in Figure~\ref{fig:trimer1}, the open trimer phase space is completely filled with regular orbits. Therefore, the Lyapunov exponents for both non-interacting systems are less than or equal to zero for any of the initial conditions.

\begin{figure}[ht]
  \centering
  \subfloat{\includegraphics[width=.45\columnwidth]{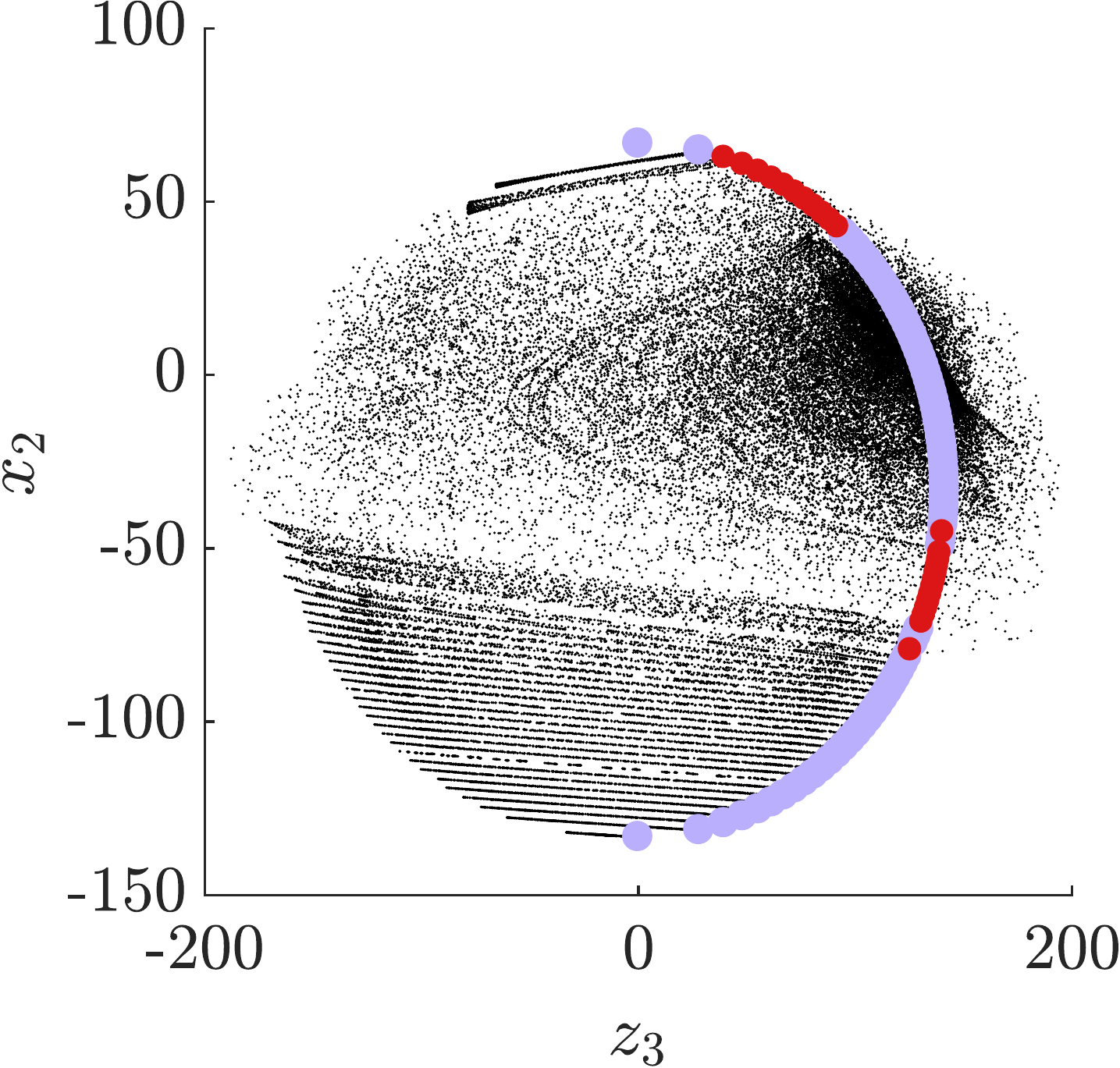}}

  \subfloat{\includegraphics[width=.45\columnwidth]{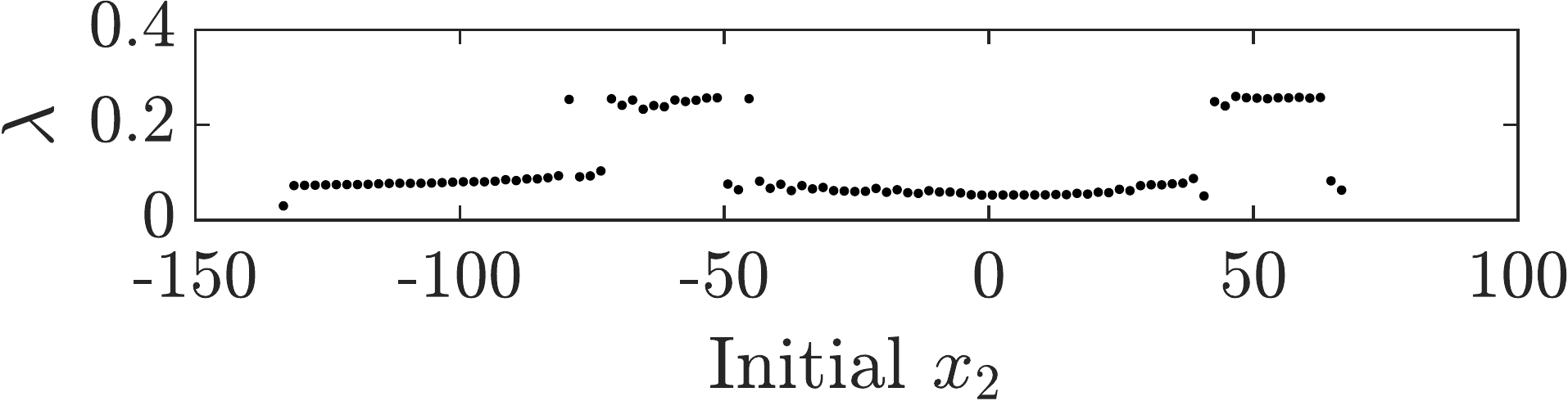}}
  \caption{(Colour online) Stroboscopic projection onto $x_2$-$z_3$ subspace (above) and FTLEs (below) at $t=150$ for the trimer with $N\chi/J=12$. Larger light blue (smaller dark red) markers on the subspace plot correspond to low (high) regions on the FTLE plot.}
  \label{fig:trimer2}
\end{figure}

For $N\chi/J \neq 0$, the equations are nonlinear and, for asymmetric initial conditions, nonintegrable.
In this case the stroboscopic projections display both regular and fuzzy regions, with the latter indicating chaotic dynamics. We have determined the utility of FTLEs by comparing their predictions to those of the stroboscopic projections in both regions. In Fig.~\ref{fig:trimer2}, we show the results of the two methods side by side, for $N\chi/J=12$. The dark red dots on the stroboscopic projections show where the FTLEs were around 0.25 in the right hand figure, and correspond well with the fuzzy chaotic regions. The bright blue dots are from the lower values of the FTLEs and correspond well with the regular regions of the stroboscopic projections. In Fig.~\ref{fig:lyaptime}, we see that there is a natural division of the FTLEs into two distinct regions, one regular and one chaotic.

\begin{figure}[ht]
  \centering
  \subfloat{\includegraphics[width=.45\columnwidth]{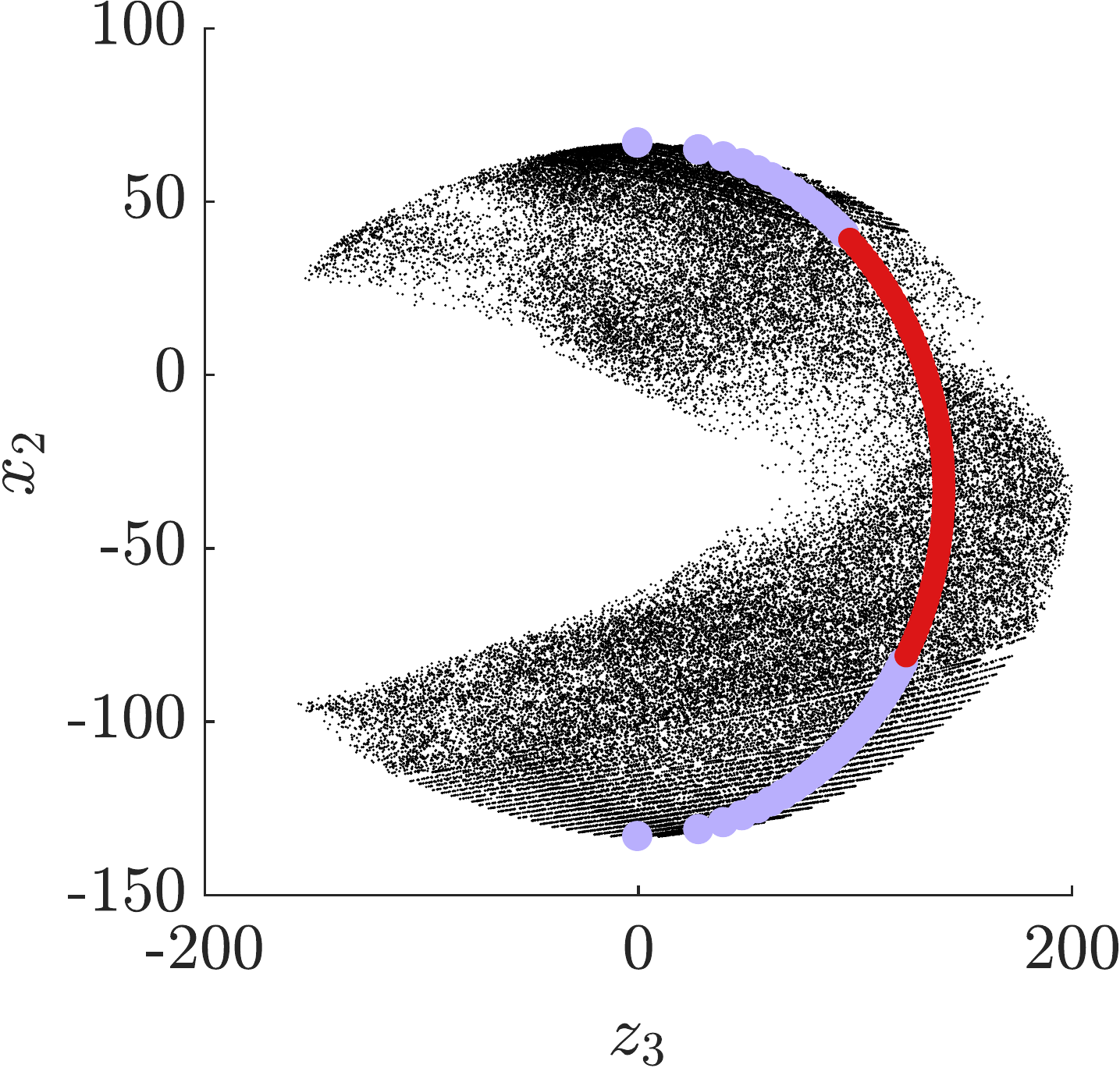}}

  \subfloat{\includegraphics[width=.45\columnwidth]{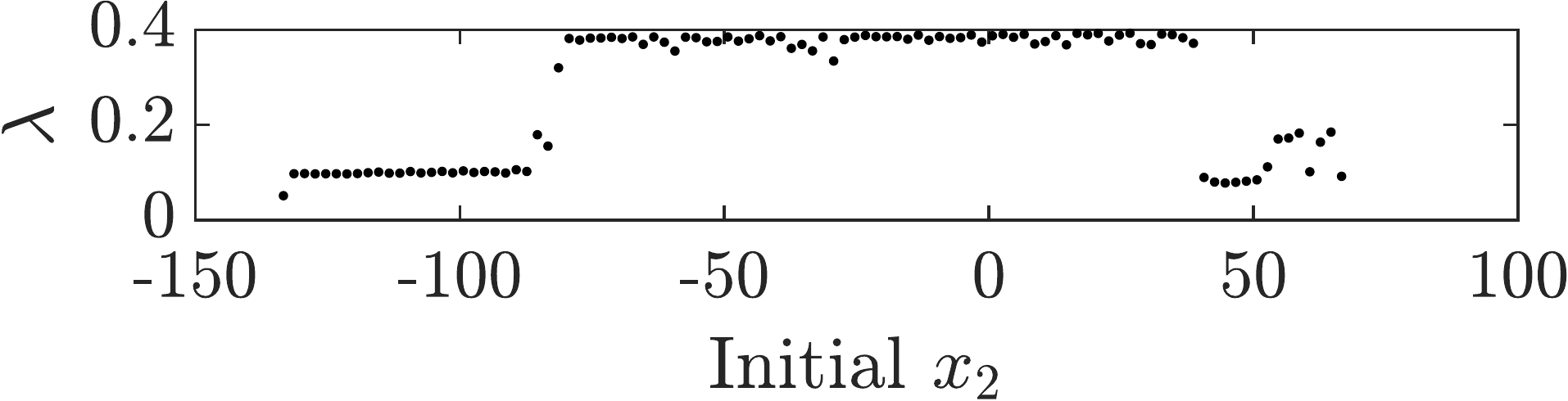}}
  \caption{(Colour online) Stroboscopic projection onto $x_2$-$z_3$ subspace (above) and FTLEs (below) at $t=100$ for the trimer with $N\chi/J=-6$. Larger light blue (smaller dark red) markers on the subspace plot correspond to low (high) regions on the FTLE plot.}
  \label{fig:trimer3}
\end{figure}

In Fig.~\ref{fig:trimer3} we show the stroboscopic section and the Lyapunov exponents for $N\chi/J = -6$. We again see regular and irregular regions in the stroboscopic section, which match well with the FTLE predictions. Although the different regions do not show up as clearly in the stroboscopic section as they did for the higher nonlinearity of Fig.~\ref{fig:trimer2}, they are still easily distinguishable.

\begin{figure}[ht]
  \centering
  \subfloat{\includegraphics[width=.45\columnwidth]{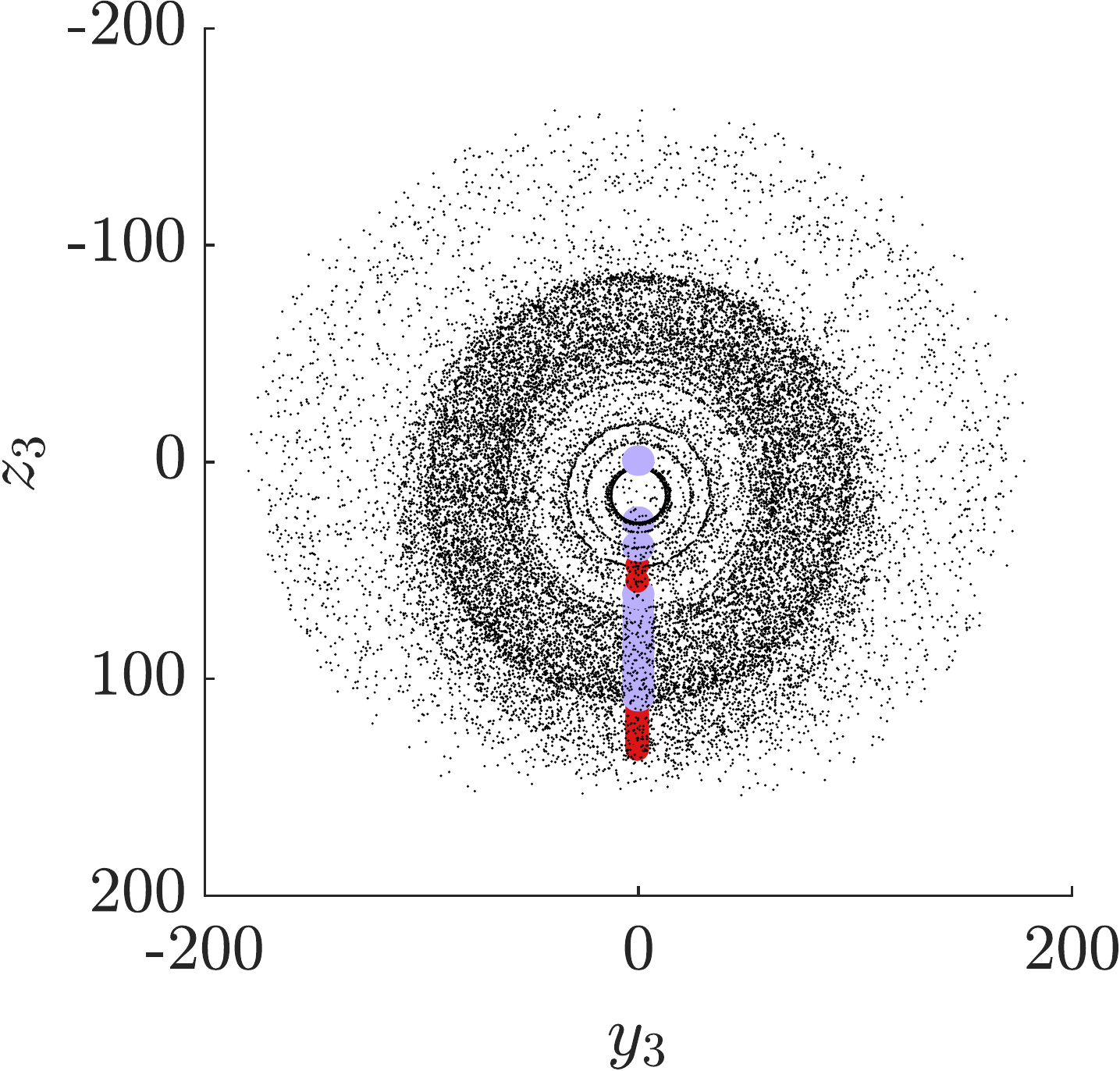}}

  \subfloat{\includegraphics[width=.45\columnwidth]{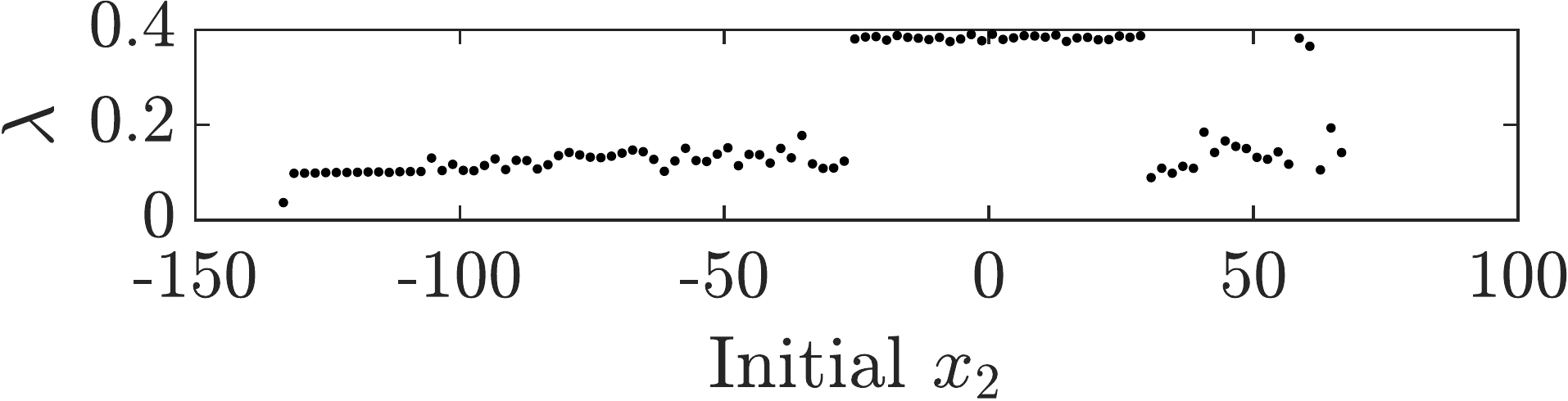}}
  \caption{(Colour online) Stroboscopic projection onto $y_3$-$z_3$ subspace (above) and FTLEs (below) at $t=100$ for the trimer with $N\chi/J=-12$. Larger light blue (smaller dark red) markers on the subspace plot correspond to low (high) regions on the FTLE plot.}
  \label{fig:trimer4}
\end{figure}

When we consider a larger negative nonlinearity, $N\chi/J = -12$, we see some interesting features, as shown in Fig.~\ref{fig:trimer4}.
Phase-space structures resembling invariant tori are present around two fixed points at high and low initial number differences, $x_2$, surrounded by a chaotic sea. The slow transition between the invariant tori and the surrounding chaotic sea is a characteristic of soft chaos, where the invariant tori cluster about KAM islands~\cite{Wimberger2014,Gutzwiller1990}. In the open trimer with a positive collisional nonlinearity, similar phenomena have previously been predicted by Mossman~\cite{Mossman2006}.

Although not shown graphically here, we found that both stroboscopic sections and FTLEs predicted that the chaotic sea will grow as the magnitude of the nonlinearity is increased, with more clearly defined tori appearing after a certain critical value of $|N\chi/J|$ between 6 and 12, where the chaotic sea begins to decrease. This result is consistent with Vardi's quantum predictions~\cite{Tikhonenkov2013}, where the open trimer only exhibited chaos for intermediate collisional nonlinearities. The diminishing chaos at large $|N\chi/J|$ can be seen in Figs.~\ref{fig:trimer2} and \ref{fig:trimer3}, where there are less high-valued FTLEs for $|N\chi/J| = 12$ than for $|N\chi/J| = 6$.

The chaotic sea in both the open trimer and quatramer systems initially grows with increasing magnitude of nonlinearity. After a certain critical value, the trimer chaotic sea begins to shrink with increasing magnitude of nonlinearity and more clearly defined tori appear for high magnitude $x_2$ and $x_3$. This result is consistent with Vardi~\cite{Tikhonenkov2013}, who considered the level spacings of the energy eigenstates in a quantum analysis. This agreement indicates that the semiclassical approximation is useful in describing the onset of open trimer chaos.

According to Ref.~\cite{Buonsante2003}, there are zero, two or four fixed points in the open trimer phase space for varying system parameters. Two such fixed points are clearly evident in Fig.~\ref{fig:trimer2} in the form of the guiding centres for the upper and lower KAM islands. The open trimer fixed point analysis undertaken in Ref.~\cite{Buonsante2003} did not consider systems with negative nonlinearity as we have here.

\subsection{The open quatramer}
\label{subsec:quatro}

\begin{figure}[ht]
  \centering
  \subfloat{\includegraphics[width=.45\columnwidth]{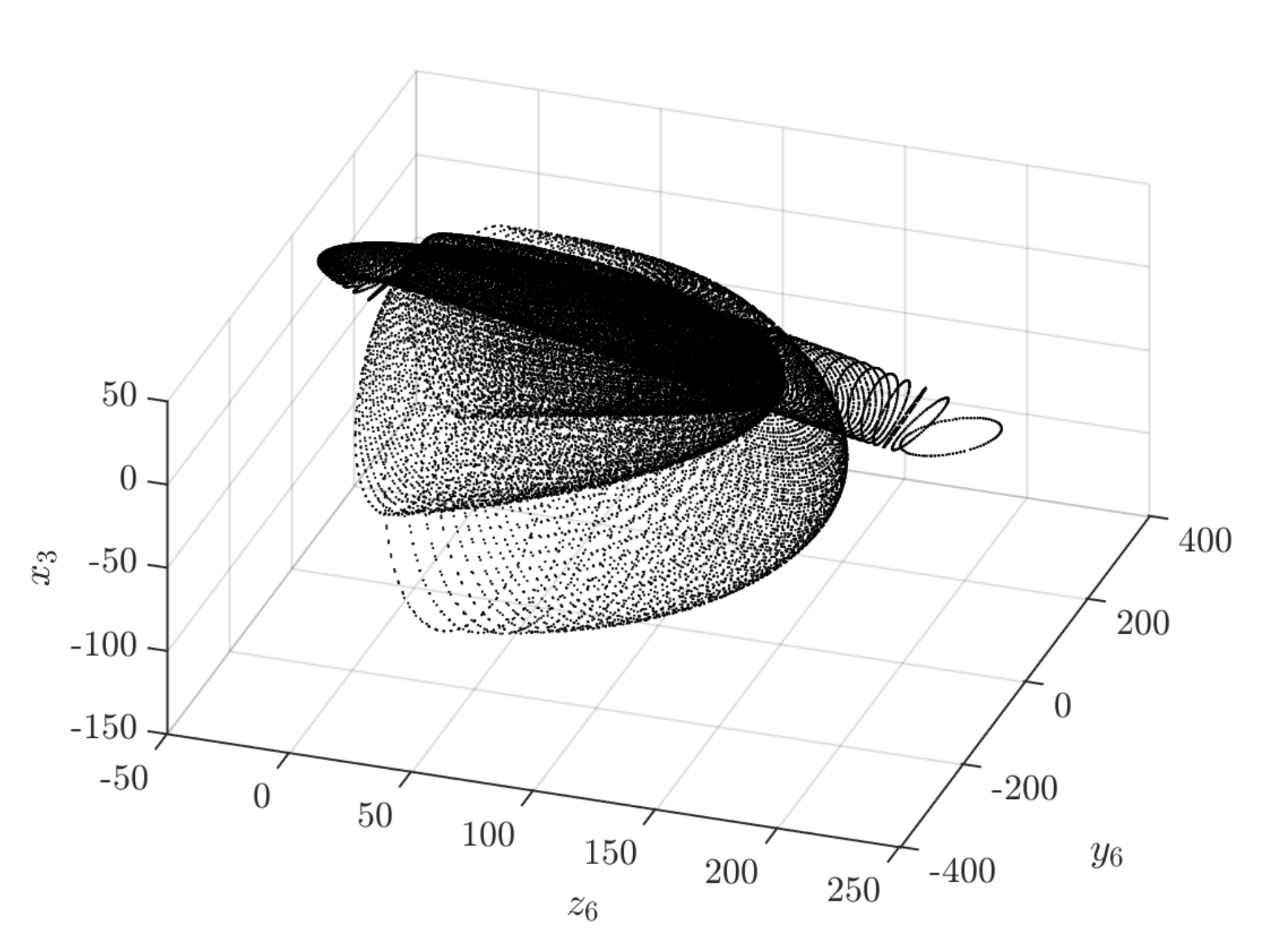}}

  \subfloat{\includegraphics[width=.45\columnwidth]{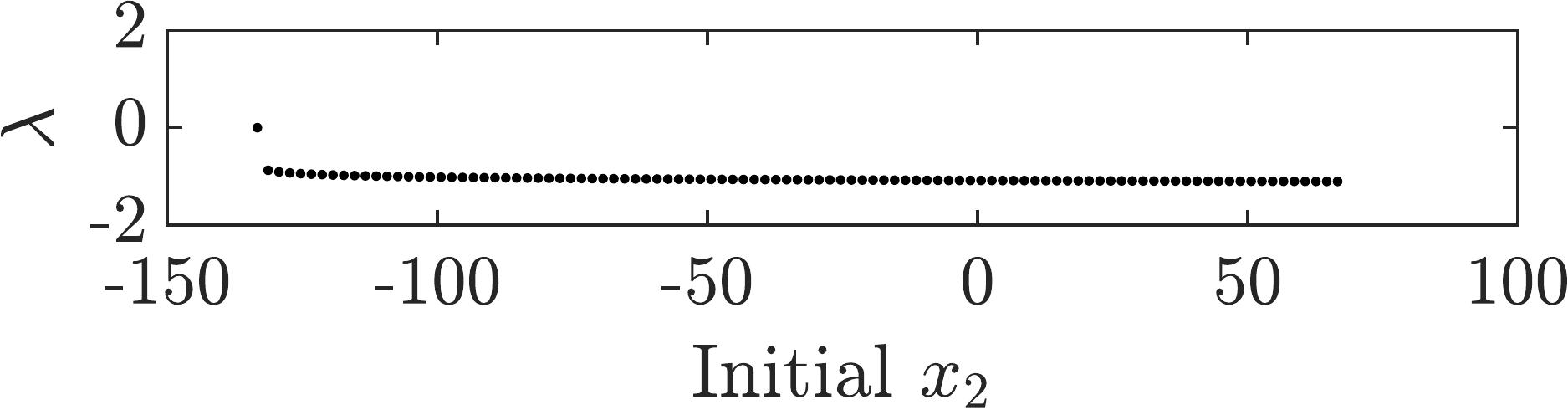}}
  \caption{Regular 3D quatramer phase space section (above) for $N\chi/J=0$ and FTLEs (below) at $t=20$.}
  \label{fig:quatramer1}
\end{figure}

The open quatramer consists of four wells in an inline configuration. Again with $N\chi/J = 0$ we find regular behaviour in the stroboscopic section and small FTLEs, as shown in Fig.~\ref{fig:quatramer1}. With $N\chi/J = 18$, we find that the behaviour is very similar to that of the trimer, as can be seen by comparing Fig.~\ref{fig:trimer2} and Fig.~\ref{fig:quatramer1}.
In each case, there is a clear division between small and large FTLEs, which also correspond well with the stroboscopic sections.

\begin{figure}[ht]
  \centering
  \subfloat{\includegraphics[width=.45\columnwidth]{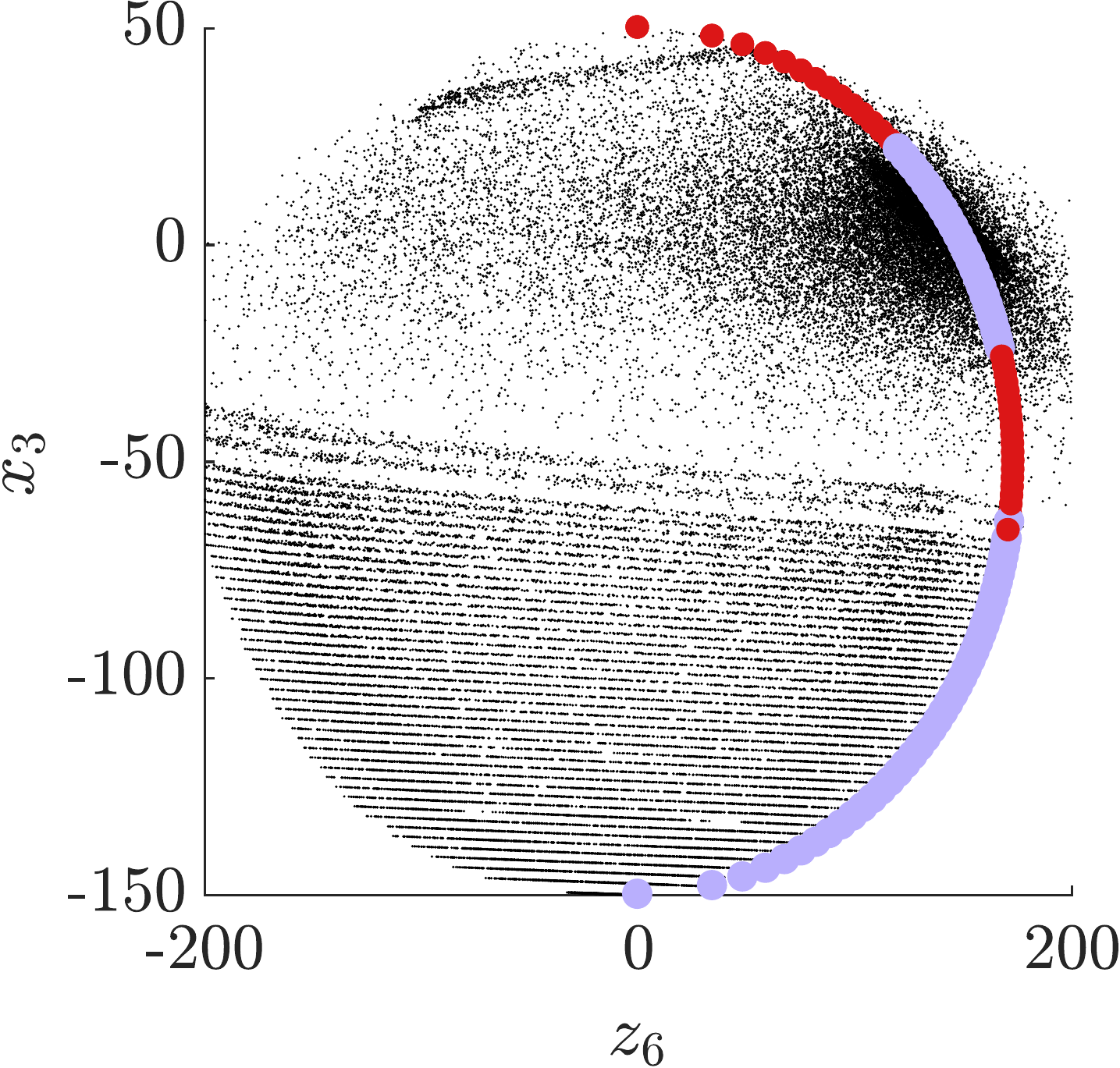}}

  \subfloat{\includegraphics[width=.45\columnwidth]{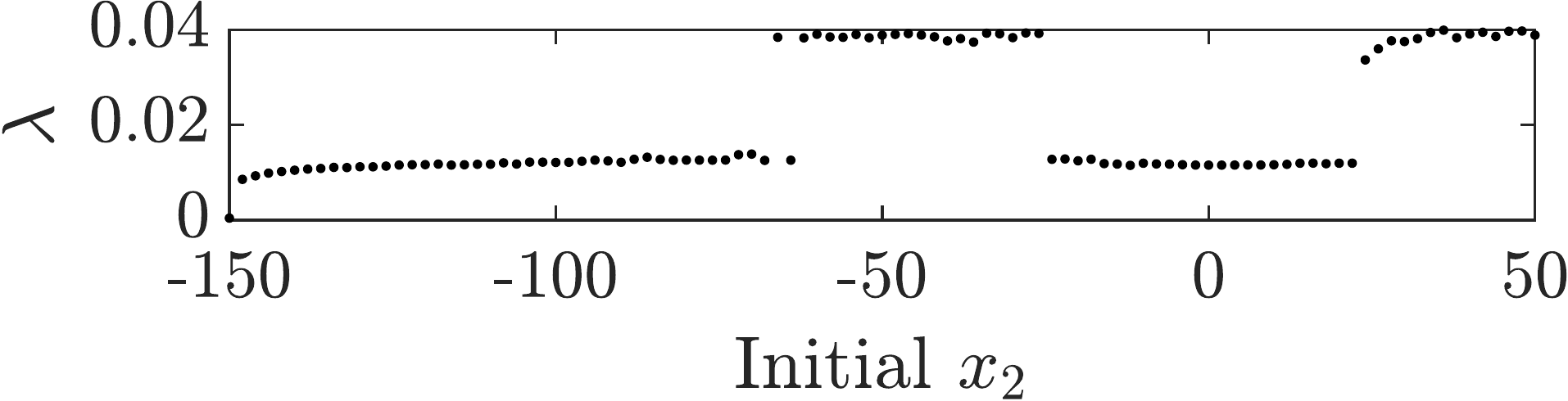}}
  \caption{(Colour online) Stroboscopic projection onto $x_3$-$z_6$ subspace (above) and FTLEs (below) at $t=1000$ for the quatramer with $N\chi/J=18$. Larger light blue (smaller dark red) markers on the subspace plot correspond to low (high) regions on the FTLE plot.}
  \label{fig:quatramer2}
\end{figure}

\begin{figure}[ht]
  \centering
  \subfloat{\includegraphics[width=.45\columnwidth]{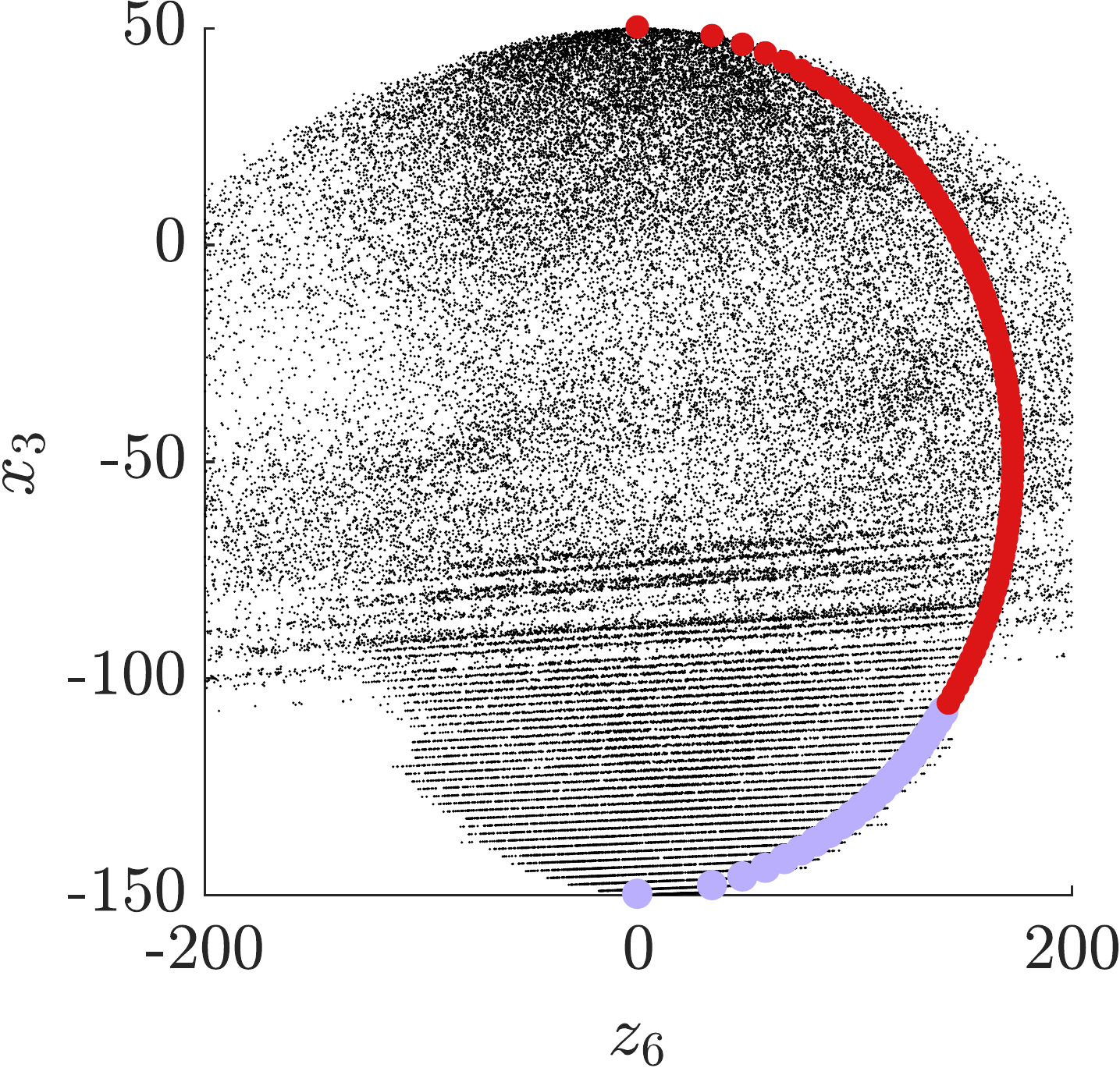}}

  \subfloat{\includegraphics[width=.45\columnwidth]{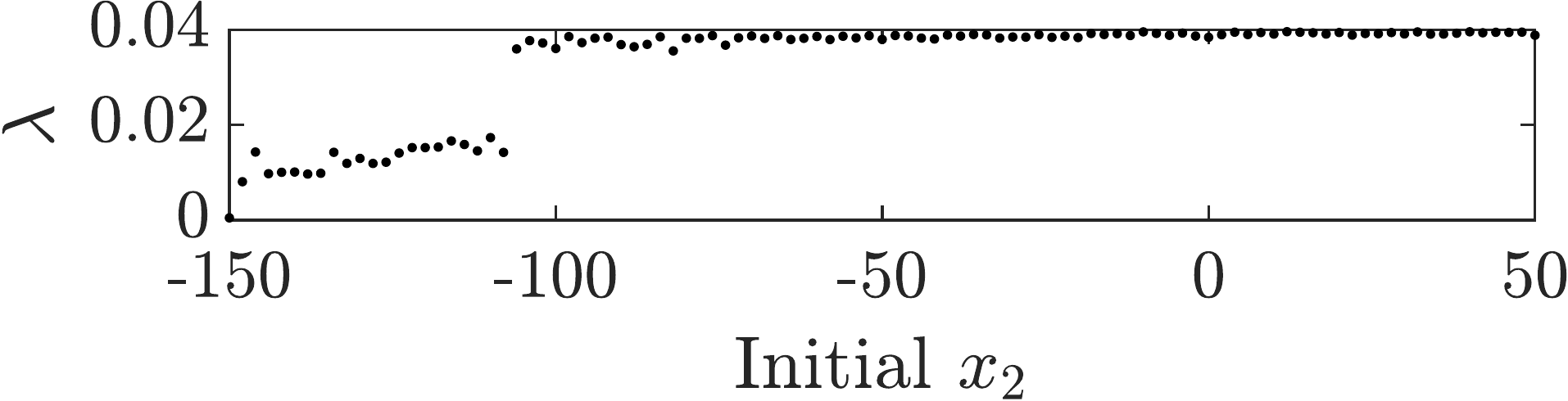}}
  \caption{(Colour online) Stroboscopic projection onto $x_3$-$z_6$ subspace (above) and FTLEs (below) at $t=1000$ for the quatramer with $N\chi/J=-18$. Larger light blue (smaller dark red) markers on the subspace plot correspond to low (high) regions on the FTLE plot.}
  \label{fig:quatramer3}
\end{figure}

However, when we investigate the effects of a negative nonlinearity, we find significant differences, as can be seen by comparison of Figs.~\ref{fig:quatramer2} and \ref{fig:quatramer3}. These differences are more pronounced for positive initial values of $x_3$.
One notable feature is that the self-trapping invariant tori that appear for the trimer at positive $x_2$ are absent from the quatramer for positive $x_3$. This indicates that for repulsive particle interactions, corresponding to positive nonlinearity, the trimer and quatramer have similar phase space dynamics. For attractive interactions and an initially depleted end-well, given by large $x_2$ ($x_3$) for the trimer (quatramer), the system displays significantly different phase space dynamics.

As with the trimer, the predictions of the quatramer stroboscopic sections and the finite-time Lyapunov exponents are consistent.
In all cases, the large magnitude FTLEs corresponded to irregular regions of the phase spaces, indicating that high relative FTLE magnitude is associated with clearly chaotic regions.

\section{Conclusions}

In this work we have shown that FTLEs provide a computationally efficient means of analysing semiclassical chaos in long chain systems with high particle number.  This result is important for two reasons. The first is that for larger systems, the dimensionality of the phase space of the stroboscopic sections becomes unmanageable, whereas the FTLE phase space, particularly in terms of the amplitudes, only expands linearly in the number of sites and thus remains computationally efficient. The second is that the FTLE is easily calculable from the equations of motion, without out having to manipulate large matrices and calculate energy eigenvalues, for example.
In all cases, we found that there was a clear separation among the FTLEs which coincided with the invariant tori and chaotic seas in the phase space stroboscopic sections. This indicates that the finite-time Lyapunov exponent method of characterising chaos will also be valid for much larger systems, where comparison with other methods can be difficult.

\begin{acknowledgments}

This research was supported by the Australian Research Council under the Future Fellowships Program (Grant ID: FT100100515).

\end{acknowledgments}

\bibliographystyle{apsrev4-1}
\bibliography{references.bib}

\end{document}